# Numerical design of nonuniform disk-loaded waveguides

M.I. Ayzatsky[1], V.V.Mytrochenko

National Science Center Kharkov Institute of Physics and Technology (NSC KIPT),
610108, Kharkov, Ukraine

On the base of modified mode matching method we obtain some results that can be useful in the process of tuning of nonunifrom disk-loaded structures. Our consideration has shown that there are some parameters that depend only on the geometric sizes of sells and equals known values for the needed phase distribution of the amplitudes of field expansion. Changing the geometric sized in such way that these parameters will tend to a given values, we can numerically design the structure that have the given phase distribution of these amplitudes. We consider these tuning parameters as a good instrument under simulation of inhomogeneous structures with sizes that usually used in accelerators.

## 1 Introduction

Waveguide structures (WSs) with longitudinal discontinuities are involved in many applications: Rf engineering, beams electronics, accelerator technique, etc. In some cases we need in creating given distributions of electromagnetic field along the WS by choosing geometric parameters of longitudinal discontinuities. Such necessity arises in the case of using the WSs for generation of electromagnetic energy by electron beams or for acceleration of charged particles by external electromagnetic fields. For these cases, there must be synchronism between particles and electromagnetic waves [1,2,3]. In most cases such synchronism will take place under a special phase distribution (PD) along the structure that is similar to the PD of a slow wave. Therefore, we must have possibility to develop the structures with the special PD along the structure and the suitable amplitude distribution (AD).

High brightness electron RF photoinjector sources, based on short-pulse laser excitation of a photocathode, have proven to be essential experimental instruments in beam physics, enabling many high impact applications [4,5]. For conserving the beam characteristics along the accelerator its RF accelerating structures must be "predictable", that is must have the field distribution coinciding with the simulation one. This gives opportunity to simulate the beam dynamics and find the way for conserving beam brightness. But often the exact geometrical sizes are unknown as under the process of post-tuning of the cells they become slightly uncertain. So, simulation can be made with some uncertainty, too. To avoid this uncertainty it is necessary to have possibility to conduct simulation of the post-tuning process.

So, first step in the development of slow-wave structures has to be their numerical design. Such design must to acknowledge the possibility of creating of chosen distribution of electromagnetic fields. Results of such design can be used directly in the process of production of WSs [6].

Recently, we proposed a method for post-tuning of the disk-loaded waveguides (DLW) with specified law of radii of openings in disks [7]. This method is needed for measured and simulation data. The last one can be obtained only under the numerical design.

In the case when discontinuities in the waveguide arrange into a periodic manner, the Floque theorem facilitate significantly the process of calculating the characteristics of eigen electromagnetic waves [8].

If there is no periodicity, a powerful approach of electromagnetic fields calculation is the mode matching method (MMM) with which the fields are described by a superposition of waveguide modes, and then the boundary conditions that the tangential fields in the cross-section of the waveguide structure must be continuous are imposed at the interface between different waveguide sections [9,10,11].

The finite-element methods, that are widely being developing recently (see, for example, [6]), also give the powerful possibilities for simulation of electrodynamic structures. Using these methods, production of accelerating structures without post-tuning of the cells can be realized. The damped-detuned accelerator structure for the JLC/NLC [12] was the first structure with dimensions determined directly using the parallel simulation code [6]. The computational challenge was to model the complex geometry close to machining tolerance in order to obtain accuracy for the cavity frequency of better than 0.01%. This is necessary for maintaining the cavity efficiency and to avoid post-tuning of the cells.

If geometric sizes are given, the MMM and finite-element methods give possibility to calculate the electromagnetic field distribution.

There are difficulties in conducting the numerical design of WSs with longitudinal discontinuities on the base of electromagnetic field distribution. It is practically impossible to find the necessary changes of geometrical parameters for obtaining the needed distribution. Indeed, in the nonuniform structure the needed distribution is realized by mutual compensation and interference of fields reflected from many inhomogeneities. So, it is difficult to extract the needed information from the field distribution.

For conducting the efficient numerical design of accelerator structures with arbitrary field distributions it is desirable to find some parameters that depend only on the geometric sizes of discontinuities and equals known values for the needed distribution.

When geometric parameters of longitudinal discontinuities change from cell to cell very slow there is combination of some field amplitudes that do not depend on field values in the tuned state [13,7]. Method for the DLW post-tuning with using electric field values measured in the middle of cavities is based on this opportunity [13].

In the modified MMM the fields are described in different parts of the WS by a superposition of closed cavity modes or waveguide ones [14,15]. Such modification gives possibility to obtain analytical equations for amplitudes of field expansions and introduce "the tuning parameters'. This paper contains results of searching and studying such parameters.

## 2 Characterization of detuned cells. Tuning parameters

Let's consider a cylindrical nonuniform DLW (Fig.1). We will consider only axially symmetric fields with $E_z, E_r, H_\varphi$ components. Time dependence is $\exp(-i\omega t)$.

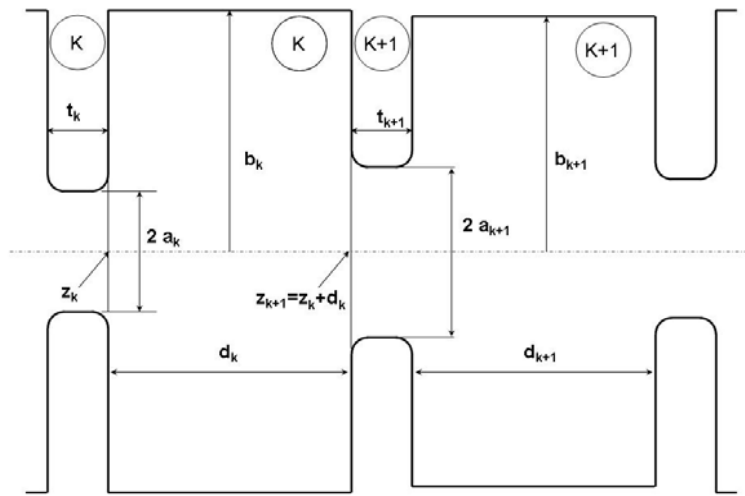

Fig. 1

In the frame of modified MMM such coupling equations can be obtained [14,15]

$$\left(\omega_{010}^{(n)2} - \omega^2\right)e_{010}^{(n)} = \omega_{010}^{(n)2}\sum_{j=-\infty}^{\infty}e_{010}^{(j)}\alpha_{010}^{(n,j)} \quad . \tag{1.1}$$

Here $e_{010}^{(n)}$ - amplitudes of $E_{010}$ modes, $\omega_{010}^{(n)}$ - eigen frequencies of these modes, $\alpha_{010}^{(n,j)}$ - real coefficients that depend on the geometric sizes of discontinuities. Sums in the right side can be truncated

$$\left(\omega_{010}^{(n)2} - \omega^2\right)e_{010}^{(n)} = \omega_{010}^{(n)2}\sum_{j=n-N}^{j=n+N}e_{010}^{(j)}\alpha_{010}^{(n,j)} \, . \tag{1.2}$$

We can rewrite (1.2) in the form

$$\sum_{s=-N}^{N}e_{010}^{(n+s)}\overline{\alpha}_{010}^{(n,n+s)} = 0 \, , \tag{1.3}$$

where

$$\overline{\alpha}_{010}^{(n,n+s)} = \left[\alpha_{010}^{(n,n+s)} - \left(1 - \frac{\omega^2}{\omega_{010}^{(n)2}}\right)\delta_{n+s,n}\right] \, . \tag{1.4}$$

Instead complex amplitudes $e_{010}^{(n)}$ we introduce real amplitudes $A_n$ and phases $\varphi_n$

$$e_{010}^{(n+s)} = A_{n+s}\exp(i\varphi_{n+s}) \, . \tag{1.5}$$

Dividing (1.3) on real and imaginary parts, we obtain

$$\sum_{s=-N}^{N}\gamma_{n,n+s}^{(1,0)}A_{n+s} = 0$$
$$\sum_{s=-N}^{N}\gamma_{n,n+s}^{(2,0)}A_{n+s} = 0 \tag{1.6}$$

where

$$\gamma_{n,n+s}^{(1,0)} = \cos(\varphi_{n+s})\overline{\alpha}_{010}^{(n,n+s)}$$
$$\gamma_{n,n+s}^{(2,0)} = \sin(\varphi_{n+s})\overline{\alpha}_{010}^{(n,n+s)} \, . \tag{1.7}$$

From Appendix 1 it follows that the systems of linear equations (1.6) have a nonzero solution if the following equalities hold

$$\gamma_{n,n+N-1}^{(1,2N-1)}\gamma_{n,n+N}^{(2,2N-1)} - \gamma_{n,n+N}^{(1,2N-1)}\gamma_{n,n+N-1}^{(2,2N-1)} = 0 \, , \tag{1.8}$$

or in more suitable for calculation form

$$\frac{\gamma_{n,n+N-1}^{(1,2N-1)}\gamma_{n,n+N}^{(2,2N-1)}}{\gamma_{n,n+N}^{(1,2N-1)}\gamma_{n,n+N-1}^{(2,2N-1)}} = 1 \, . \tag{1.9}$$

In this case the amplitude $A_{n+N}$ of the $(n+N)$-th cell depends only on the amplitude of the $(n+N-1)$-th cell

$$A_{n+N} = -\frac{\gamma_{n,n+N-1}^{(2,2N-1)}}{\gamma_{n,n+N}^{(2,2N-1)}}A_{n+N-1} = -\frac{\gamma_{n,n+N-1}^{(1,2N-1)}}{\gamma_{n,n+N}^{(1,2N-1)}}A_{n+N-1} \, , \tag{1.10}$$

From (1.18) and (1.7) it follows that fields with $\varphi_n' = -\varphi_n$ and $\varphi_n' = \varphi_n + \Delta$ will have the same amplitude distributions $A_n' = A_n$.

If we know $\gamma_{n,n+s}^{(1,0)}, \gamma_{n,n+s}^{(2,0)}$ for the cells with numbers $n_1 \le n \le n_2$, the relations (1.8) can be calculated for $n_1 \le n \le n_2 - (2N-1)$ (see (1.18)-(1.21) in Appendix 1) and we can find relations between the amplitudes in the cells from the $(n_1+N-1)$-th cell to the $(n_2-N)$-th one $(A_{n_1+N-1}, A_{n_1+N}, A_{n_1+N+1}, ... A_{n_2-N})$.

As we will see later, it is convenient to assign index $(n+2)$ to the left side of the equality (1.9)

$$G_{n+2}^{(N)} = \frac{\gamma_{n,n+N-1}^{(1,2N-1)}\gamma_{n,n+N}^{(2,2N-1)}}{\gamma_{n,n+N}^{(1,2N-1)}\gamma_{n,n+N-1}^{(2,2N-1)}} \ . \tag{1.11}$$

These parameters do not depend on the field distribution. If the PD ($\{\varphi_n\}$) is specified, $G_n^{(N)}$ depend on the geometric sizes only. Changing the geometric sizes of longitudinal discontinuities in such way that the coefficients $G_n^{(N)}$ tend to 1 is the goal of numerical design. If we can do this, the PD of waveguide structure with longitudinal discontinuities will coincide with specified one $\{\varphi_n\}$. So, we can name $G_n^{(N)}$ as "tuning parameters'.

We must emphasize that this procedure gives possibility to tune the PD of the $e_{010}^{(n)}$ amplitude. Indeed, the PD of the full field in the middle of the cells can be obtained by summing the series

$$\vec{E}^{(k)} = \sum_{m,n} e_{0mn}^{(k)} \vec{E}_{0mn}^{(k)}(r, z = z_k + d_k / 2) \ , \tag{1.12}$$

where $e_{010}^{(k)}$ are determined by expressions (1.5) and (1.10), and $e_{0mn}^{(k)}$ ($m, n \neq 1, 0$) are determined by summation of such series [14,15]

$$\left(\omega_{n,m}^{(k)2} - \omega^2\right) e_{0,n,m}^{(k)} = \sum_{j=k-N}^{j=k+N} e_{010}^{(j)} H_{n,m}^{(k,j)} \ . \tag{1.13}$$

From (1.12) and (1.13) it follows that for the nonuniform waveguide there is difference between the PD of full field and the PD of the $e_{010}^{(k)}$ amplitude. This difference is determined by the distinction between the geometries the considered waveguide and the uniform one.

### 3 Properties of the tuning parameters

To illustrate the influence of "defects" on the value of the tuning parameters let's consider the example of the uniform DLW. In this paper we will consider the DLWs with a constant phase shift per cell equals $\varphi = 2\pi / 3$.

In Fig. 2 - Fig. 3 results of calculations of the tuning parameters for the uniform DLWs ($a$=1.6 cm, $b$=4.40137 cm, $d$=1.6792 cm, $t$=0.4 cm, $r$=0.2 cm - Fig. 2; $a$=1.4 cm, $b$=4.1893 cm, $d$=3. 099 cm, $t$=0.4 cm, $r$=0.2 cm - Fig. 3) with small "defect" in the 21$^{st}$ sell are presented.

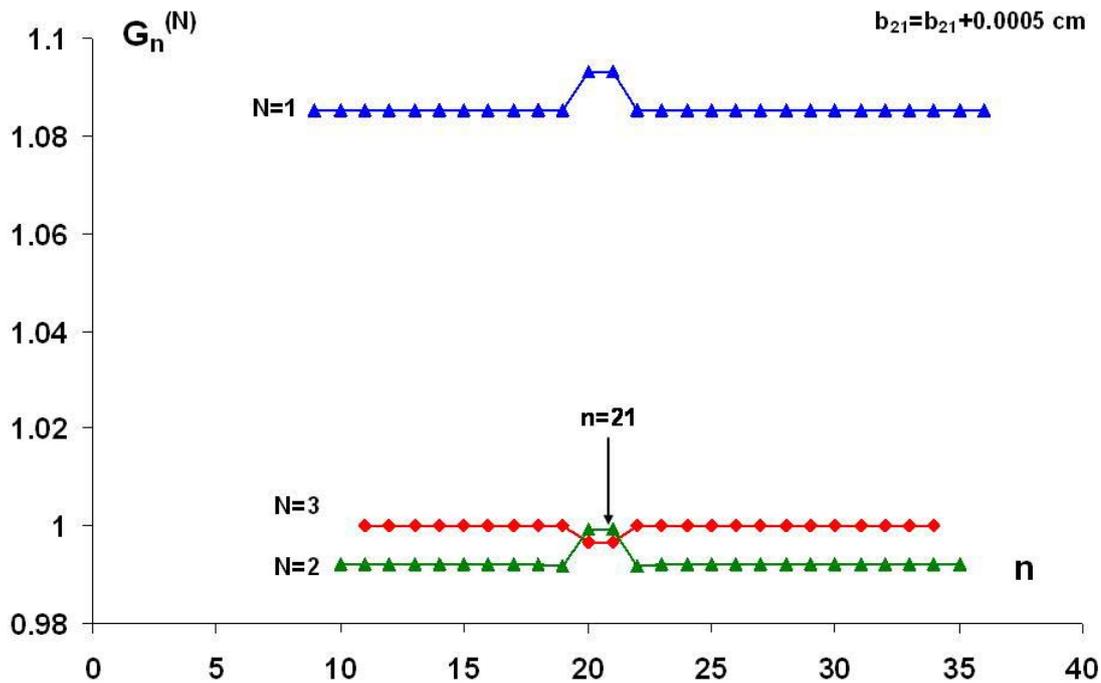

Fig. 2 Dependence of the tuning parameter on the cell number for the uniform DLW with small "defect" in the 21st sell

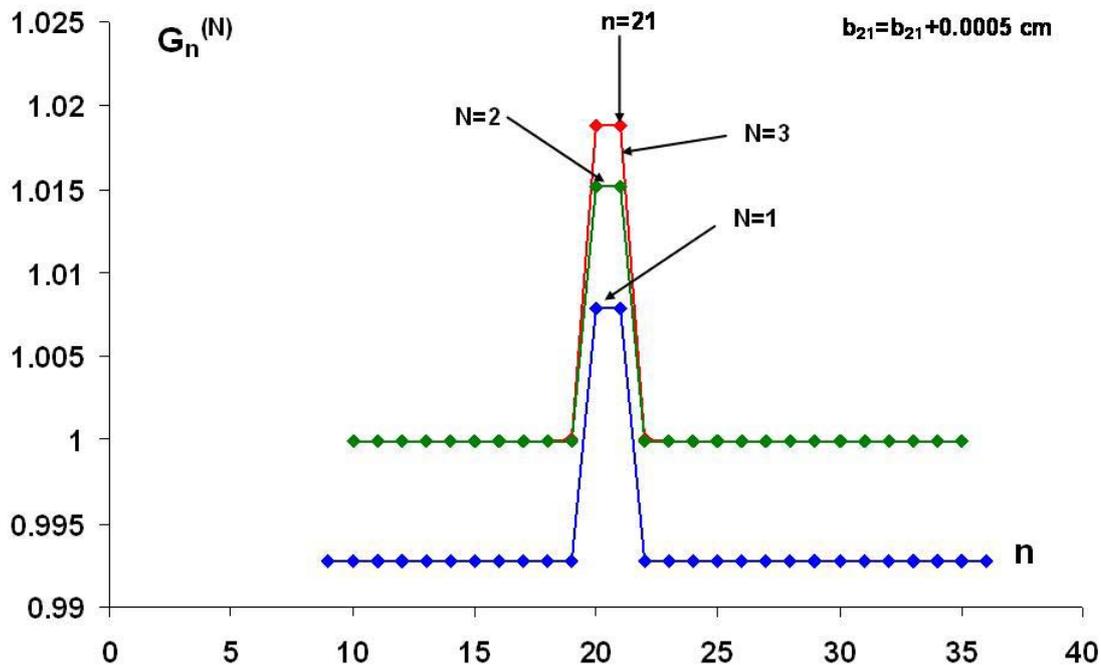

Fig. 3 Dependence of the tuning parameter on the cell number for the uniform DLW with small "defect" in the 21st sell

Results of calculations show that for the uniform DLWs with strong coupling (Fig. 2) only $G_n^{(3)}$ (seven coupling cells) take the necessary values. If the coupling becomes weaker, we can use $G_n^{(2)}$ (Fig. 3) as the tuning parameter.

From these results it follows that introduced tuning parameters $G_n^{(N)}$ can be used for characterizing the deviation the radius of the cell from the tuned one. But if you want to change the value of tuning parameter for some cell you have to change the radius of the next cell[2]. Lets note that the increment if $G_n^{(3)}$ for strong and weak coupling has a different sign.

## 4 Numerical design of nonuniform disk-loaded waveguides

On the base of coupling equations (1.2) we can study the processes of wave propagation in inhomogeneous DLWs. It can be done if we suppose that before the inhomogeneous DLW there is a homogeneous fragment of DLW and after the inhomogeneous DLW there is a biperiodic homogeneous fragment[3]. In this case in homogeneous fragments of DLW at sufficient distance[4] from the connection interfaces (when all evanescent wave decay) we can search amplitudes in the form

$$e_{010}^{(k)} = \begin{cases} \exp\{i\varphi_{1,0}(k-k_1+1)\} + R\exp\{-i\varphi_{1,0}(k-k_1+1)\}, & k < k_1 \\ T_1\exp\{i\varphi_{2,0}(k-k_2-1)\} & k = k_2+1, k_2+3,\dots , \\ T_2\exp\{i\varphi_{2,0}(k-k_2-1)\} & k = k_2+2, k_2+4,\dots \end{cases} \quad (1.14)$$

where $R$ is the reflection coefficient and $T_1, T_2$ are the "transition" coefficients.

### 4.1 SLAC type structure

One of the most popular structures is the SLAC constant gradient structure [16]. We consider the process of tuning of the front 9 cells ($d$ =2.9148 cm, $t$ =0.5842 cm, $r$ =.0 cm) without taking into account attenuation in the walls[5]. In Fig. 4-Fig. 12 results of simulation are presented. The initial values of the cell radii were taken under assumption that each cell is the part of the homogeneous DLW with disks that coincide with the left-side disk in inhomogeneous stack. The initial reflection coefficient was equal 2.58E-3. It is follows from Fig. 11 that the initial DLW has a large additional phase shift (near 1 degree per cell). In the process of consecutive cell tuning the cell radii were changed in such way that the deviations of the tuning coefficients from one become as small as possible. In our case we reached values of $G_n^{(2,3)}$ that differ from one no more than 0.00015 (see Fig. 7). Under such tuning process we reached deviations of phase shift per cell from $2\pi/3$ less than $\pm0.015$ degree for the phase of the $e_{010}^{(n)}$ amplitudes and phase of the longitudinal electric field in the middle of the cells (see Fig. 12). The reflection coefficient after tuning equals 1.2E-4. Note that under tuning process the cell radii of the nonuniform part (cells N 19-27) were changed to about 4 microns (see Fig. 5). For DLW under consideration after the nonuniform part we obtained the biperiodic structure that is needed to fulfil relations (1.9).

---

[2] We remind that we consider the case $\varphi = 2\pi/3$

[3] Investigation shows that usually after the nonuniform part we must arrange the biperiodic DLW to fulfil relations (1.9)

[4] We must remember that the homogeneous fragments must contain more then 6 resonators.

[5] Tuning with taking into account attenuation will be considered in the future work

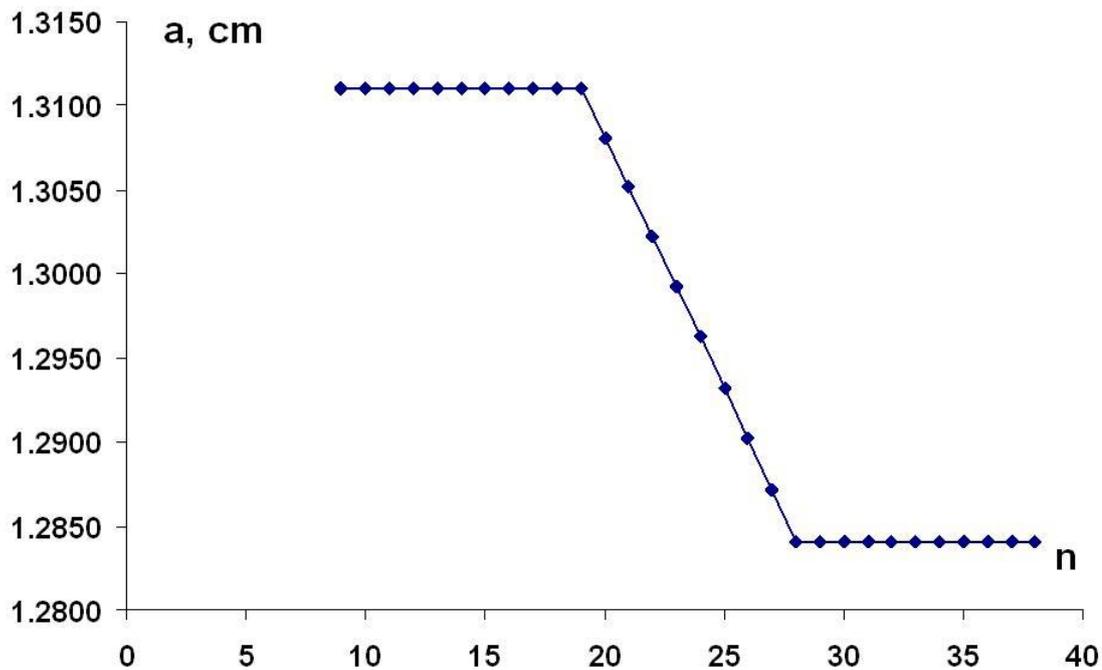

Fig. 4 Radius of disk openings as a function of the cell number

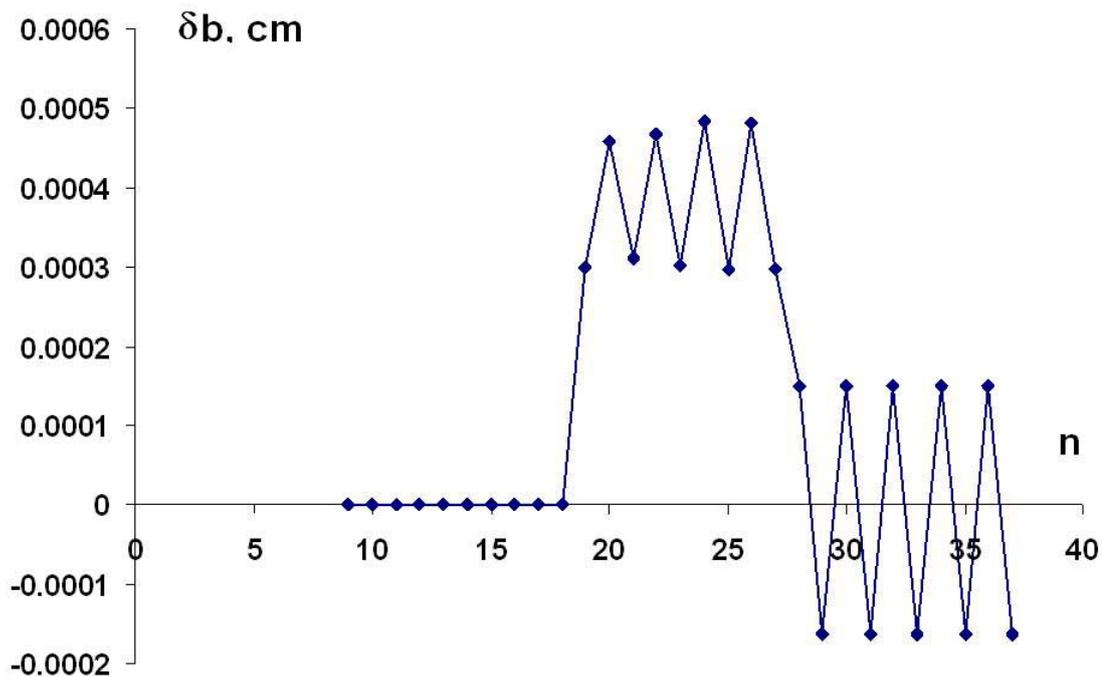

Fig. 5 Deviation of cavity radius ($b_n^{(before)} - b_n^{(after)}$) after tuning process as a function of the cell number

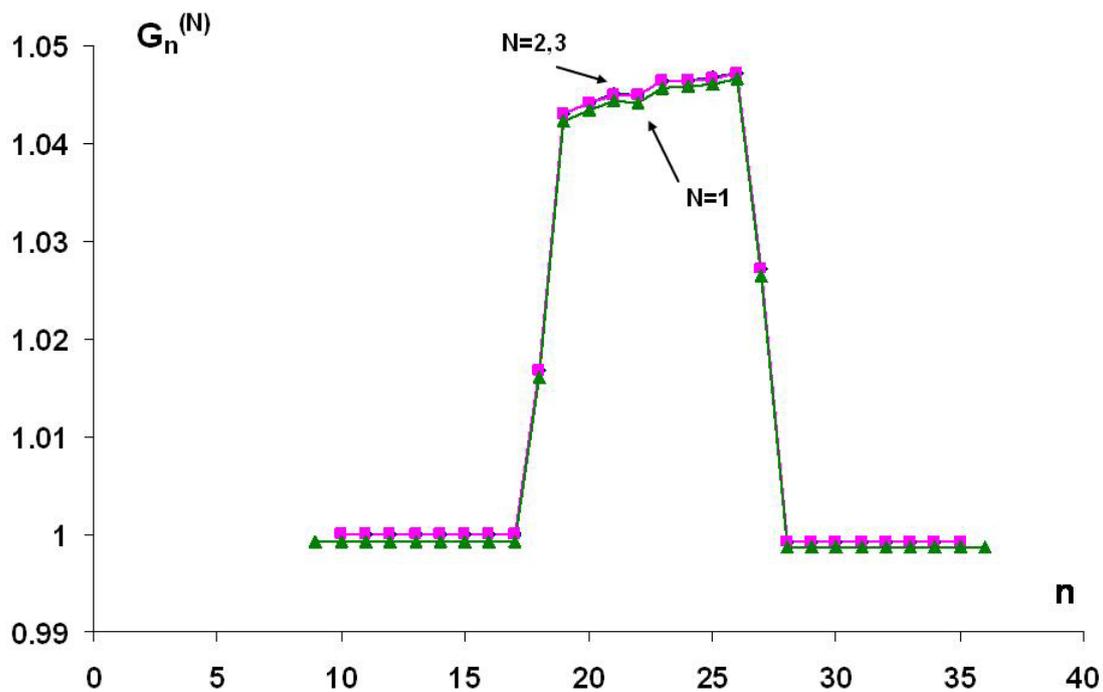

Fig. 6 Tuning parameter as a function of the cell number before tuning for different number of coupling cells

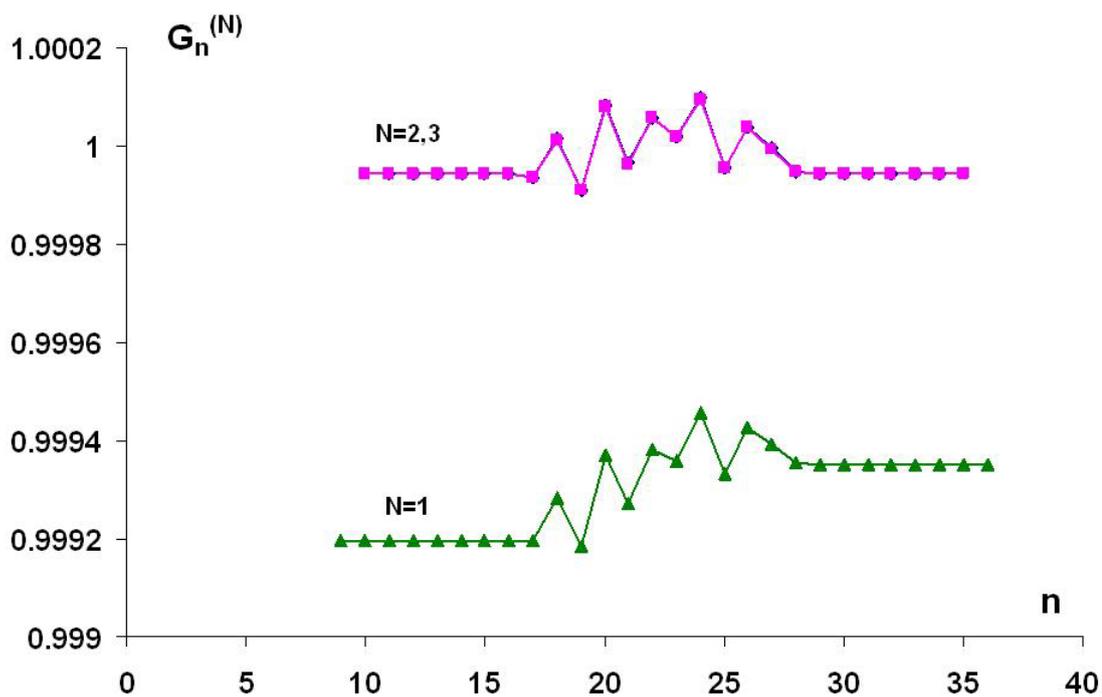

Fig. 7 Tuning parameter as a function of the cell number after tuning for different number of coupling cells

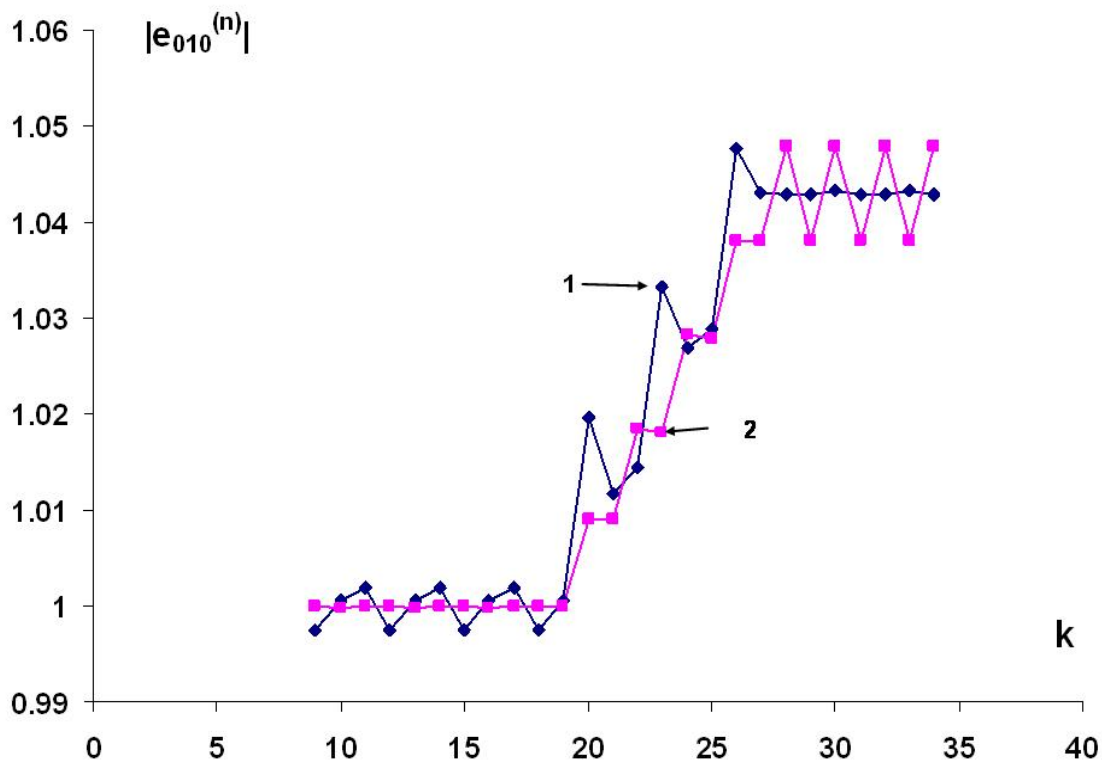

Fig. 8 Module of the $e_{010}^{(n)}$ amplitude as a function of the cell number before (1) and after (2) tuning

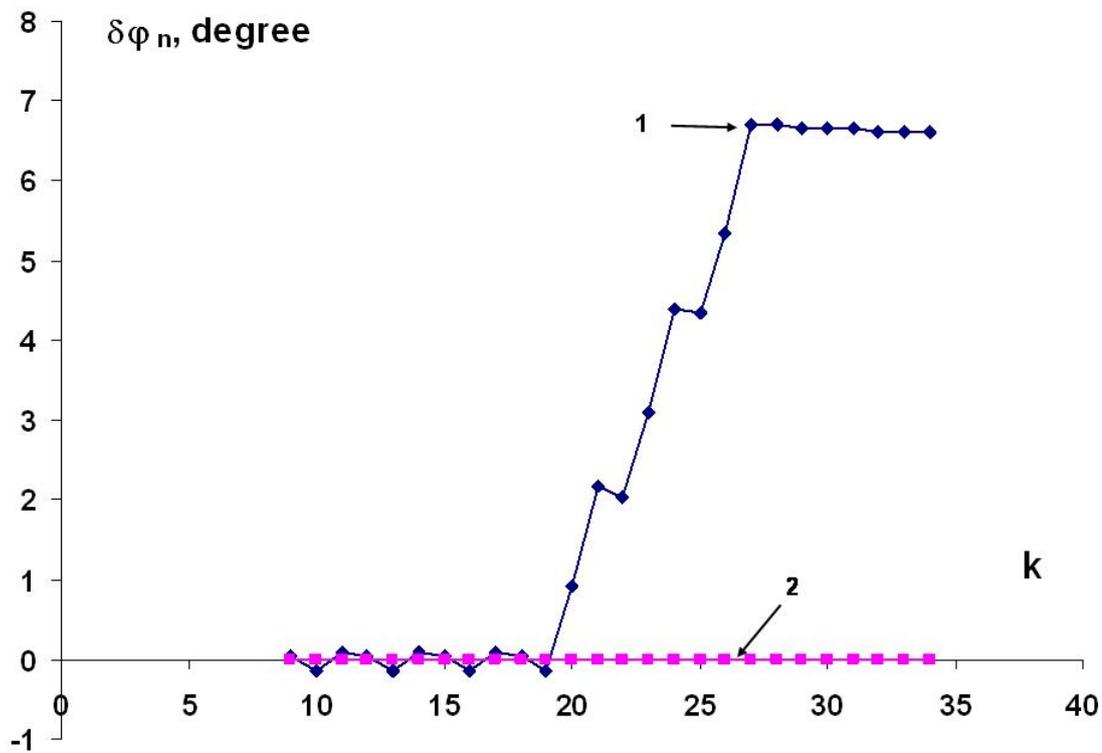

Fig. 9 Phase of the $e_{010}^{(n)}$ amplitude as a function of the cell number before (1) and after (2) tuning

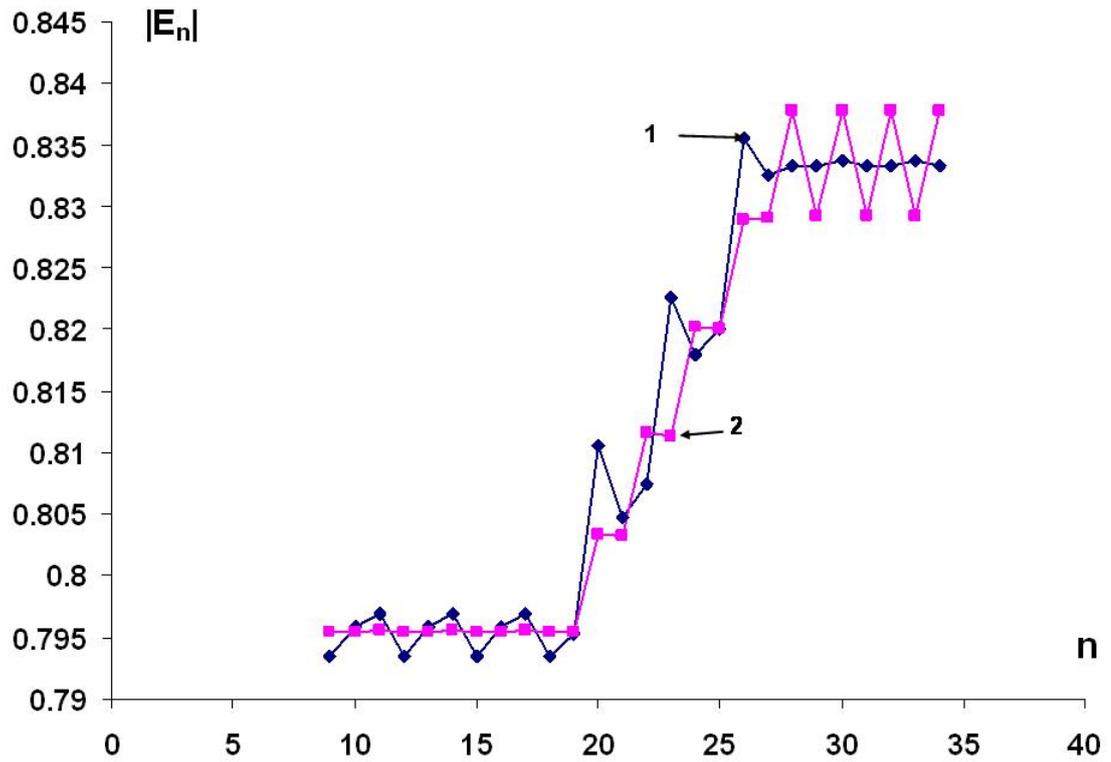

Fig. 10 Module of the longitudinal electric field in the middle of the cell as a function of the cell number before (1) and after (2) tuning

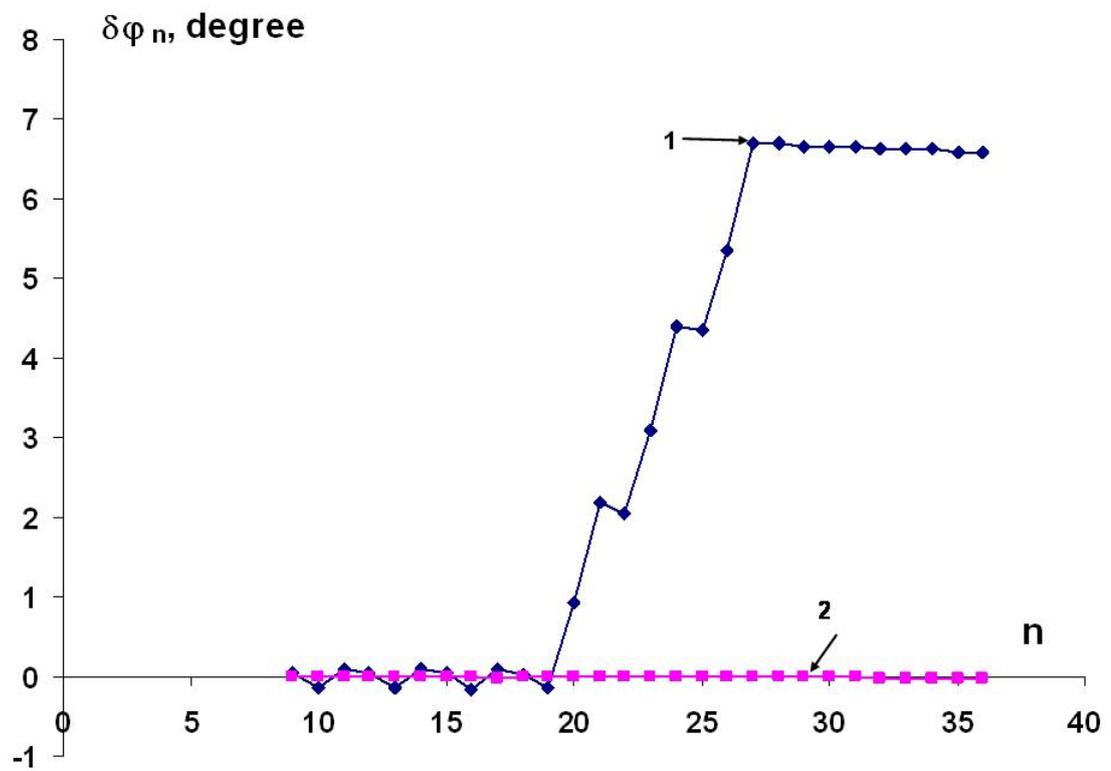

Fig. 11 Phase of the longitudinal electric field in the middle of the cell as a function of the cell number before (1) and after (2) tuning

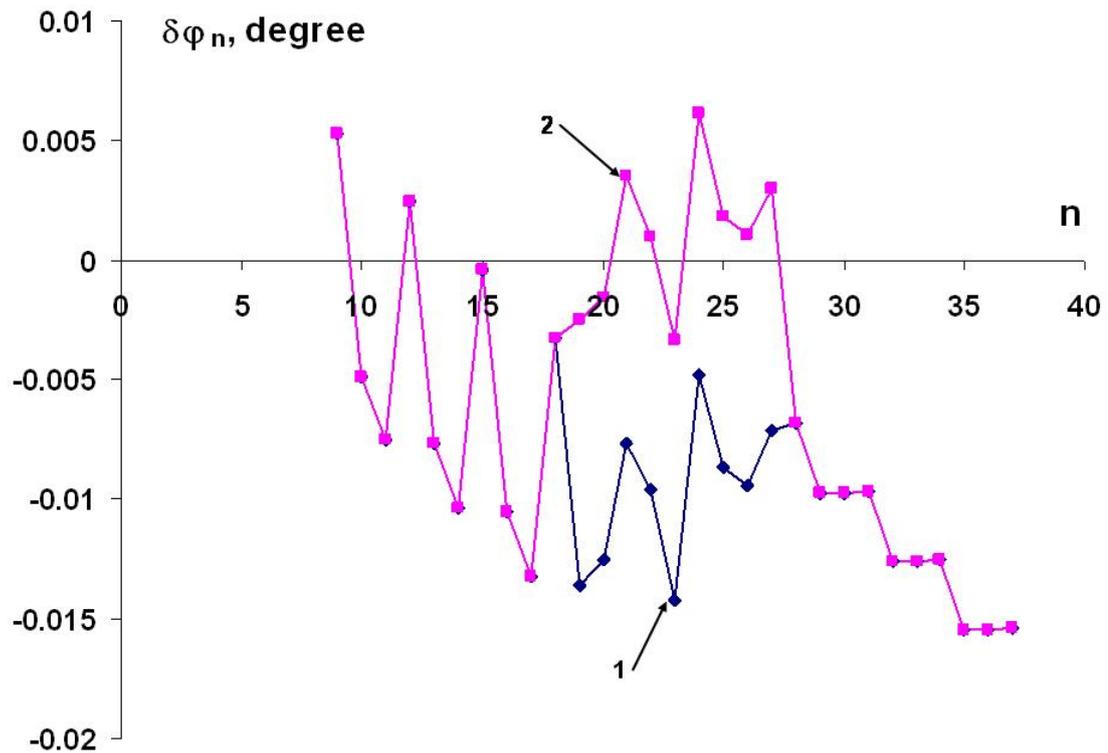

Fig. 12 Phase of the $e_{010}^{(n)}$ amplitudes (1) and phase of the longitudinal electric field in the middle of the cells (2) as a function of the cell number after tuning

### 4.2 Connection of two different uniform DLWs

In paper [7] we considered the connection of two different uniform DLWs with using one transition cell. Transition cell sizes were chosen by making the reflection coefficient small enough. We showed that it is impossible to choose the transition cell sizes that give the phase distribution without phase skip. We tried to tune such connection using the proposed above method.

Results of of the tuning process for the connection of two DLWs with such sizes $a_I$ =1.6 cm, $b_I$ =4.36528 cm, $d_I$ =1.6792 cm, $t_I$ =0.4 cm, $r_I$ =0 cm; $a_{II}$ =1.4 cm, $b_{II}$ =4.16874 cm, $d_{II}$ =3. 099 cm, $t_{II}$ =0.4 cm, $r_{II}$ =0 cm are presented in Fig. 13- .Fig. 16. The two DLWs are connected via diaphragm with the opening radius of the second DLW. The value of reflection coefficient after tuning is 6.82E-4. From Fig. 15 it follows that we can obtain very small deviation from $2\pi/3$ shift for the phase of the $e_{010}^{(n)}$ amplitudes. But there is significant deviation of the phase of the longitudinal electric field in the middle of one cell (Fig. 16). It is explained by the big differences in the geometry of two DLWs (see above). We also obtained the biperiodic structure after conection with large oscillations of field amplitudes (Fig. 14).

Simple coupling model on the base of the quasi-static approach [17] shows that characteristics of connection under consideration can be improved by using the intermediate disk with the opening radius equals $a_T = \sqrt{a_I a_{II}}$. Results of the tuning process for connection with such disk are presented in Fig. 17-Fig. 20. The value of reflection coefficient after tuning is 4.84E-4.We can see that oscillations of field amplitudes after the connection decreased significantly, but deviation of the phase of the longitudinal electric field in the middle of one cell decreased only by factor two.

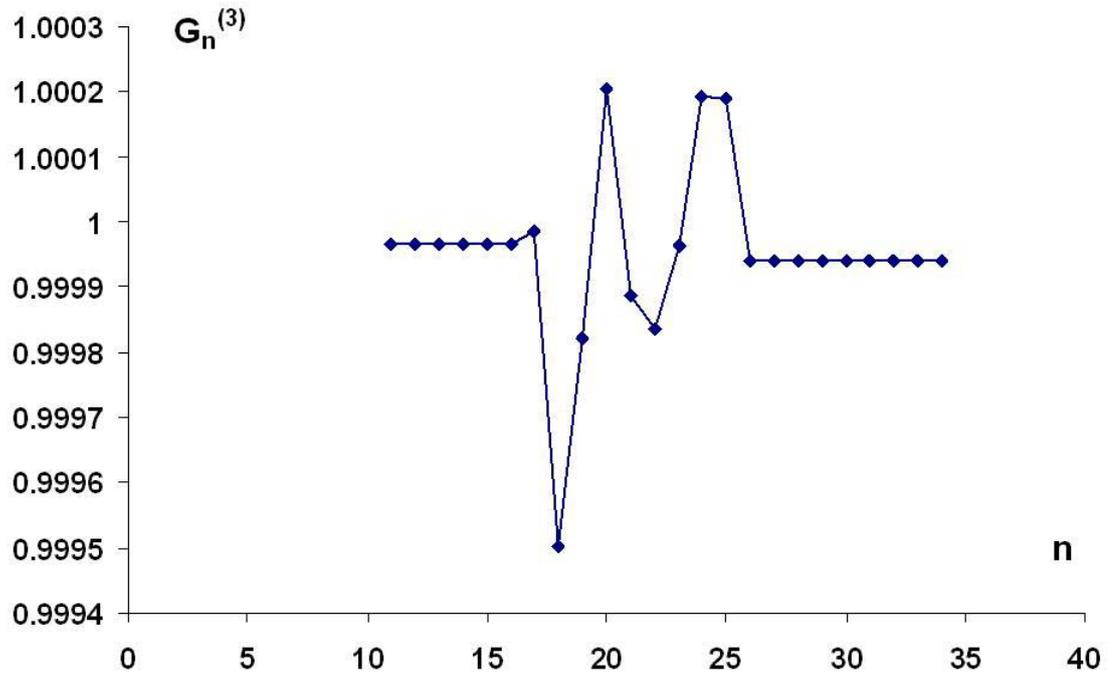

Fig. 13 Tuning parameter as a function of the cell number after tuning the connection of two different uniform DLWs without intermediate disk

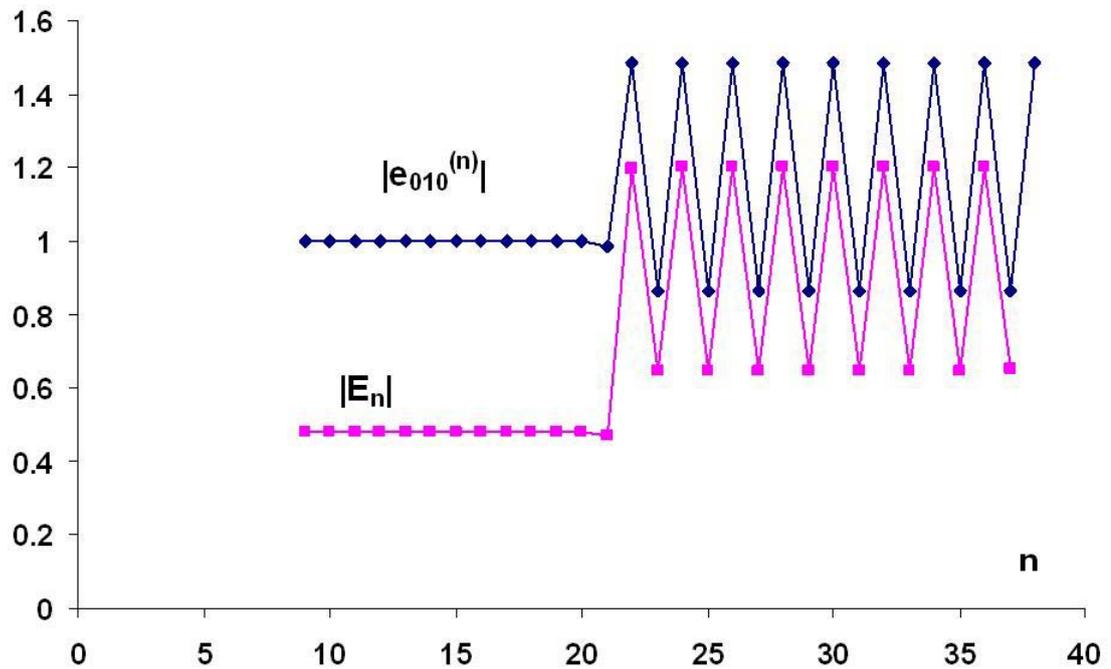

Fig. 14 Module of the $e_{010}^{(n)}$ amplitude and module of the longitudinal electric field in the middle of the cell as a function of the cell number after tuning. Intermediate disk is absent.

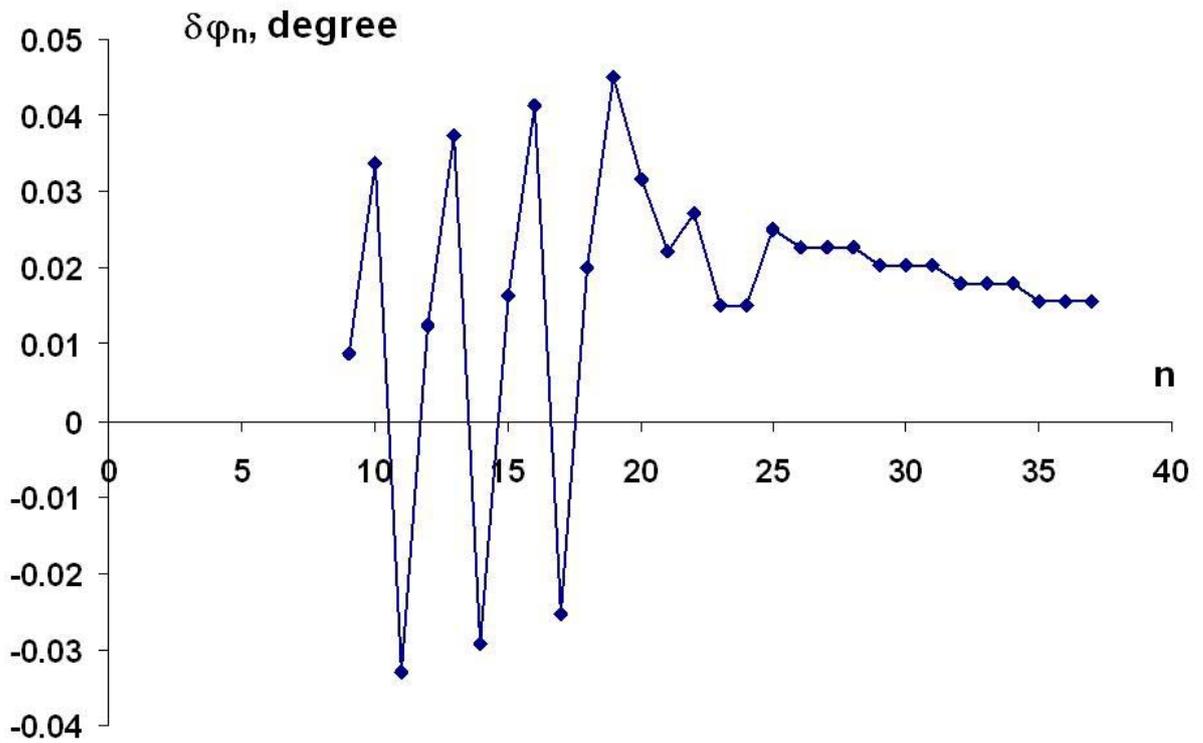

Fig. 15 Phase of the $e_{010}^{(n)}$ amplitudes as a function of the cell number after tuning. Intermediate disk is absent.

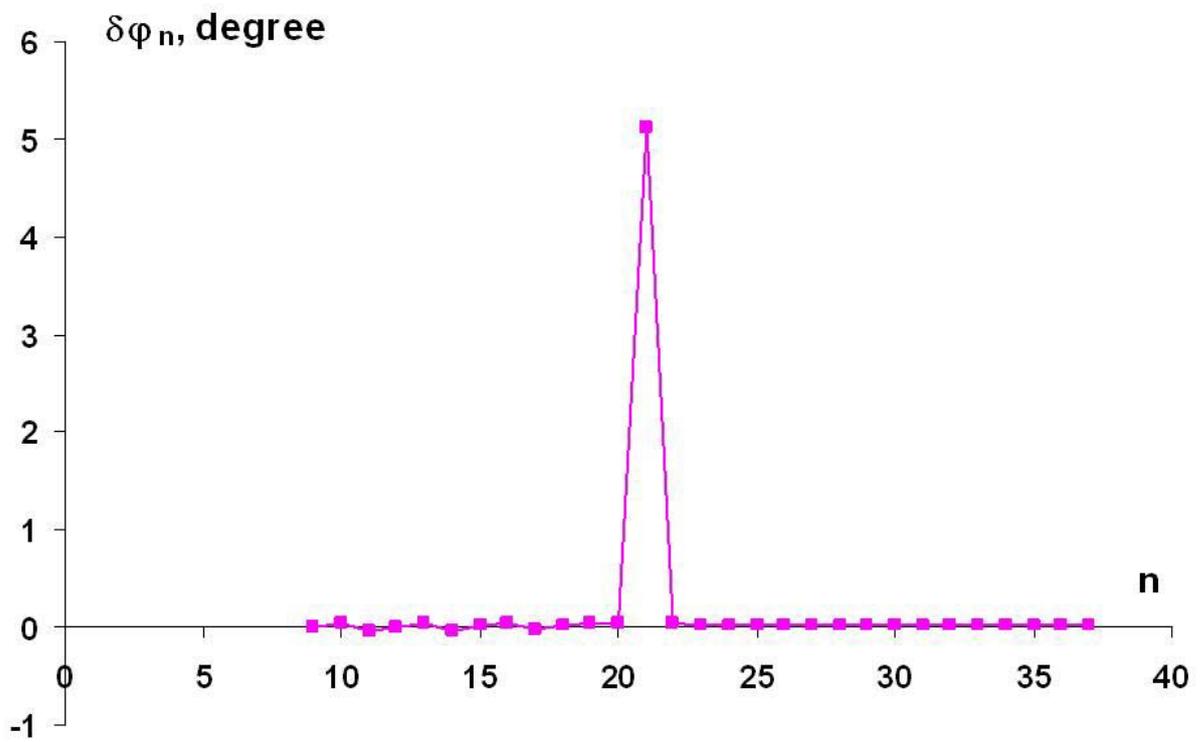

Fig. 16 Phase of the longitudinal electric field in the middle of the cells as a function of the cell number after tuning. Intermediate disk is absent.

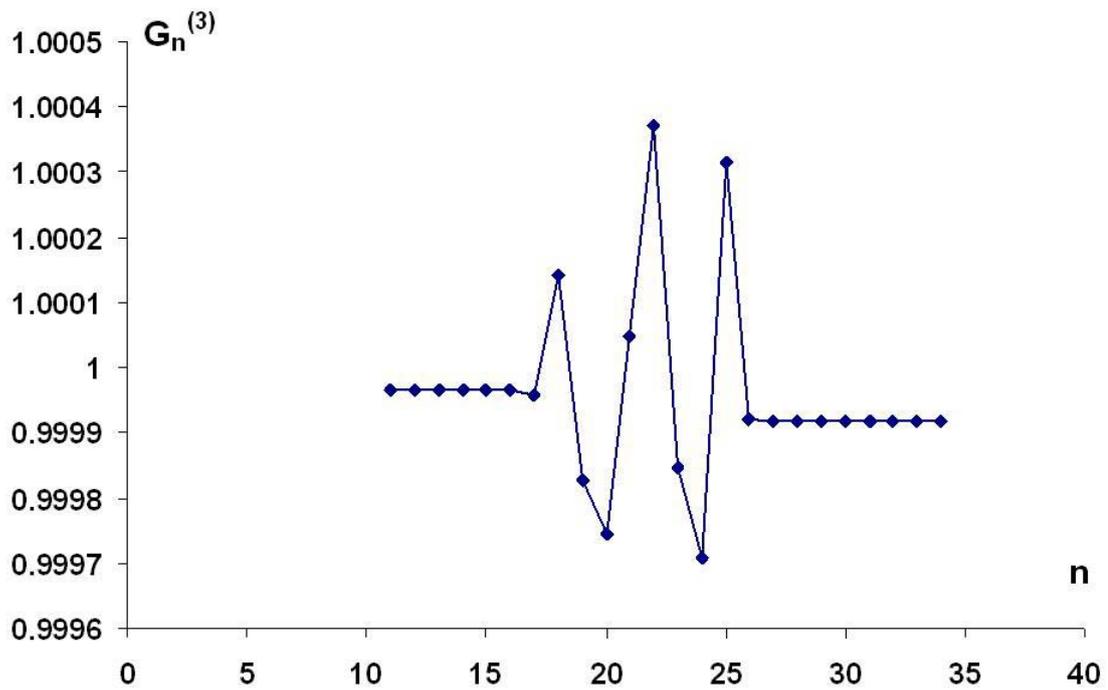

Fig. 17 Tuning parameter as a function of the cell number after tuning the connection of two different uniform DLWs with intermediate disk

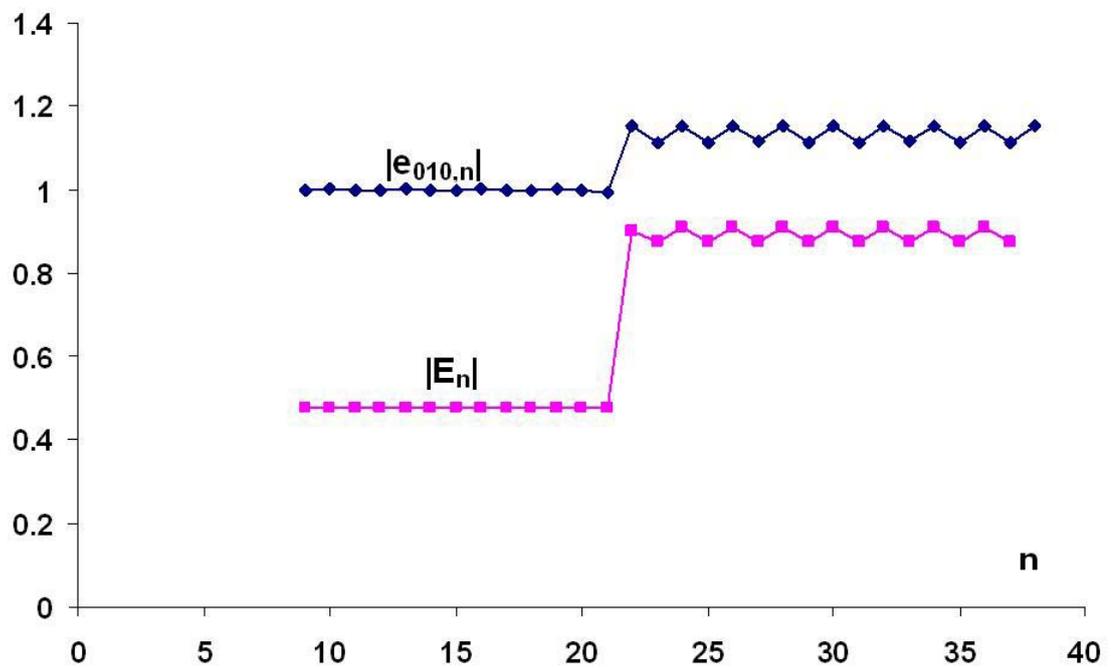

Fig. 18 Module of the $e_{010}^{(n)}$ amplitude and module of the longitudinal electric field in the middle of the cell as a function of the cell number after tuning. Intermediate disk is present.

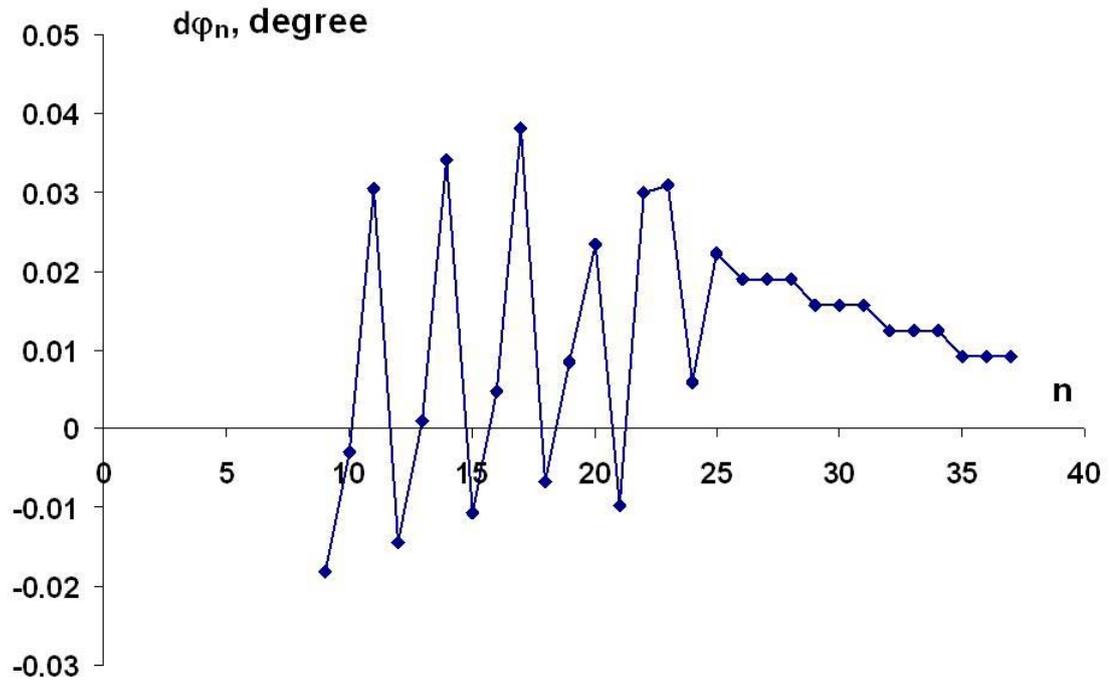

Fig. 19 Phase of the $e_{010}^{(n)}$ amplitudes as a function of the cell number after tuning. Intermediate disk is present.

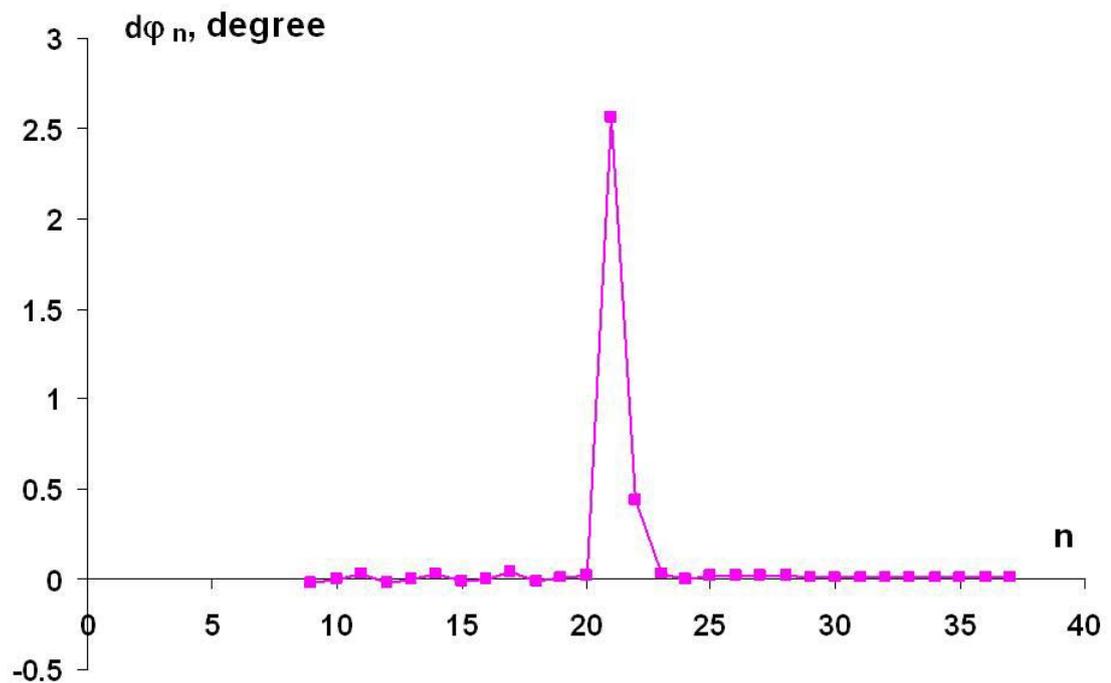

Fig. 20 Phase of the longitudinal electric field in the middle of the cells as a function of the cell number after tuning. Intermediate disk is present.

### 4.3 Front part of the injector section

An accelerating section for industrial high power linac [18], that is constructing at NSC KIPT, is the DLW with phase advance of $2\pi/3$ and variable phase velocity from 0.6c to c. It has variation in opening radii and in cavity lengths .We made numerical tuning process of the front part of this section (Fig. 21, $t = 0.4$ cm, $r = 0.2$ cm) with using introduced above tuning parameters. Results are presented in Fig. 22-Fig. 25. We can see that there are small deviations of the phase of the longitudinal electric field in the middle of several cells (Fig. 25). But they are small and will not affect strongly on the beam dynamics.

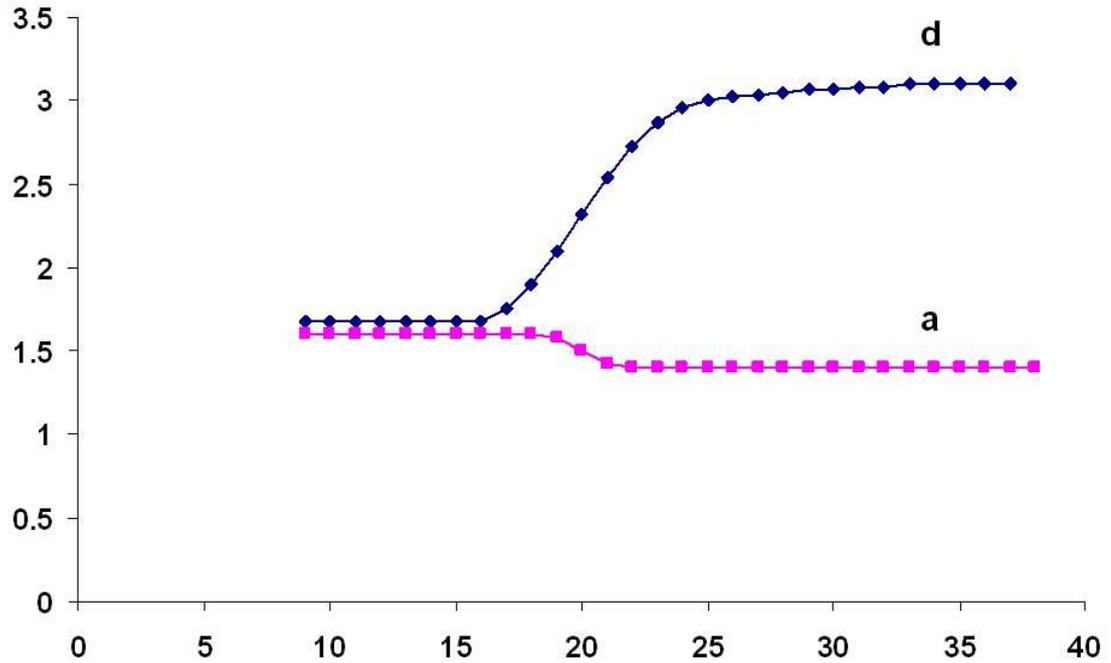

Fig. 21 Radii of openings in the disk and lengths of the cells of front part of the injector section

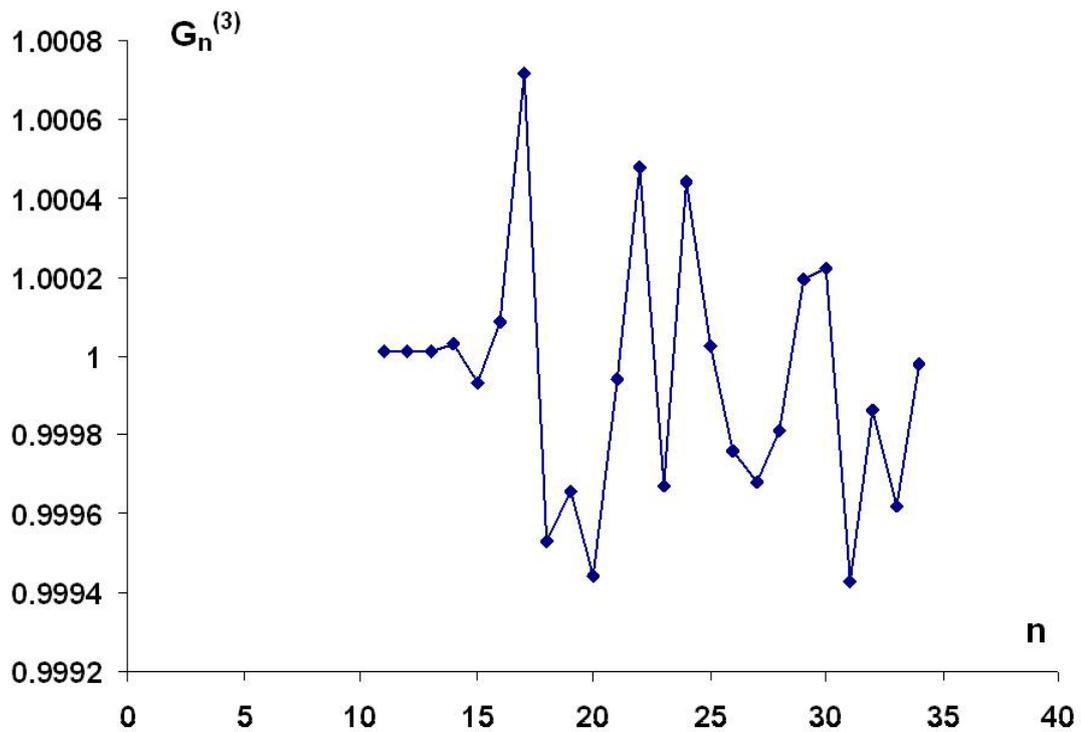

Fig. 22 Tuning parameter as a function of the cell number after tuning the front part of the injector section

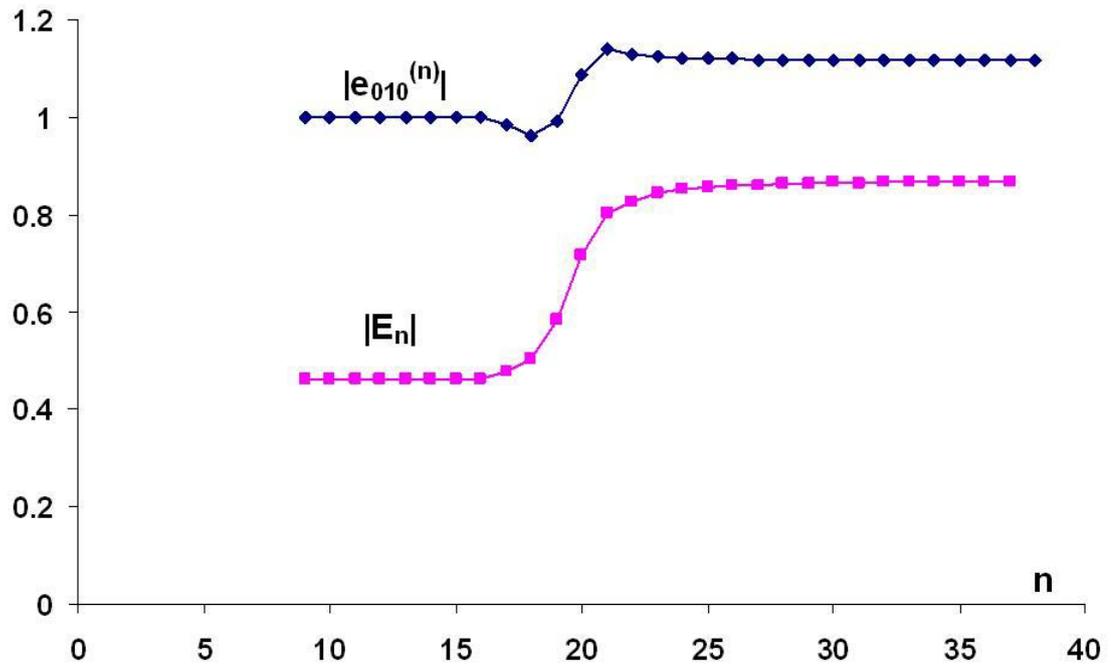

Fig. 23 Module of the $e_{010}^{(n)}$ amplitude and module of the longitudinal electric field in the middle of the cell as a function of the cell number after tuning the front part of the injector section

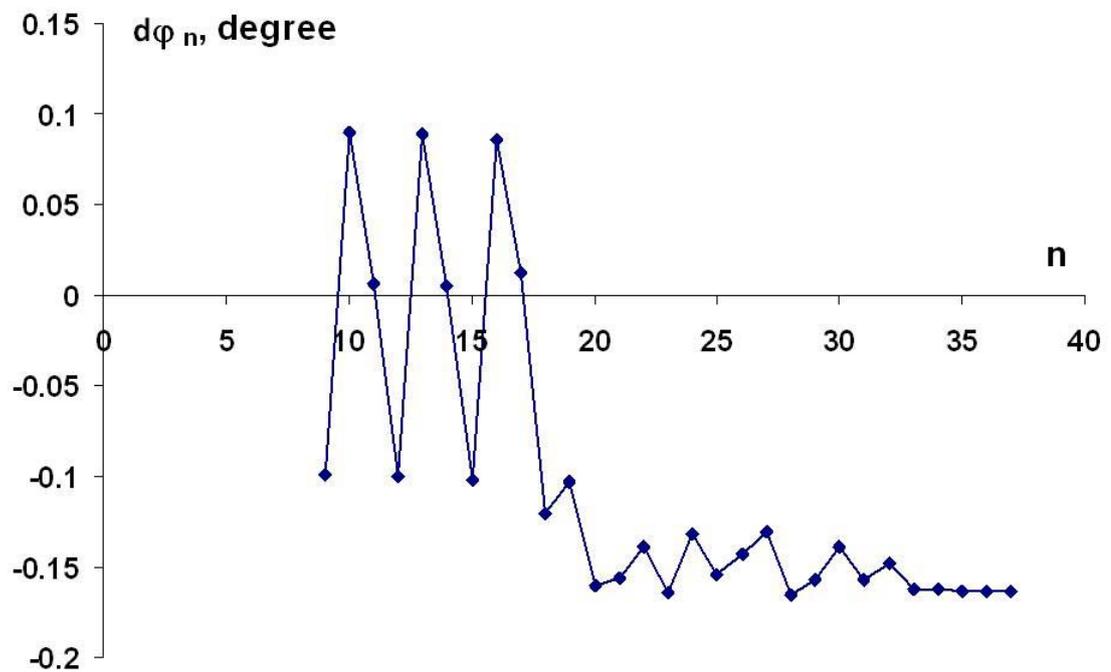

Fig. 24 Phase of the $e_{010}^{(n)}$ amplitudes as a function of the cell number after tuning the front part of the injector section.

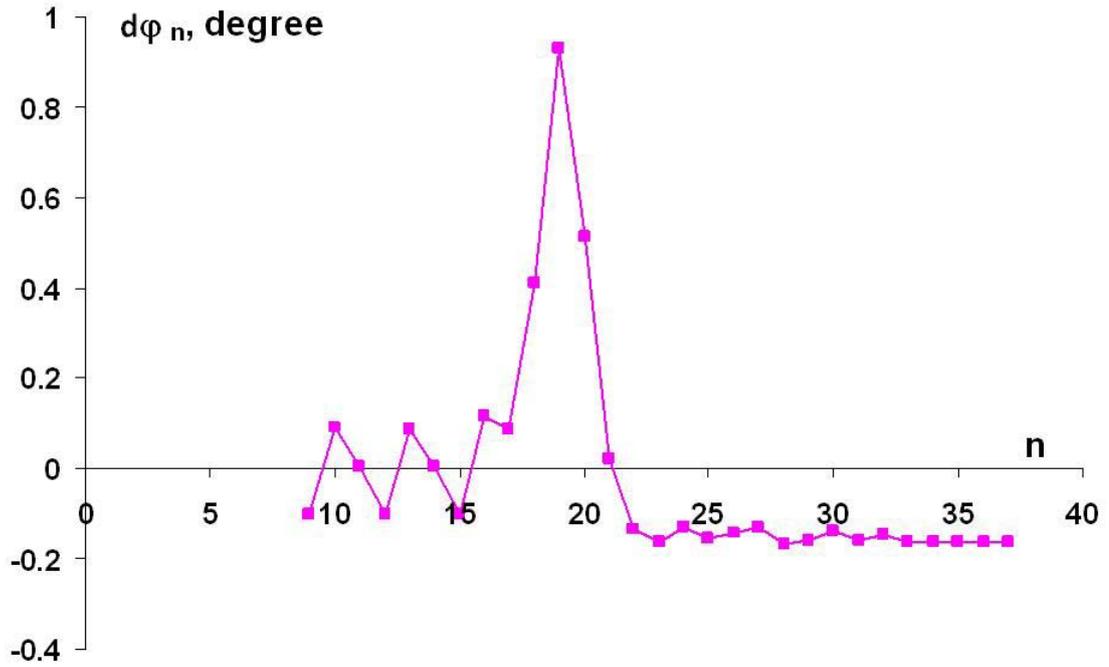

Fig. 25 Phase of the longitudinal electric field in the middle of the cells as a function of the cell number after tuning the front part of the injector section.

## Conclusions

On the base of the modified MMM we obtain some results that can be useful in the process of tuning of nonunifrom DLWs. Our consideration has shown that there are some parameters that depend only on the geometric sizes of discontinuities and equals known values for the needed phase distribution of the $e_{010}^{(n)}$ amplitude. Changing the geometric sized in such way that these parameters will tend to a given values, we can numerically design the structure that have the given phase distribution of the $e_{010}^{(k)}$ amplitude. Despite that for the nonuniform waveguide there is difference between the PD of full field and the PD of the $e_{010}^{(k)}$ amplitude, this differences can be small and tuning parameters will be good instrument under simulation of inhomogeneous structures.

## Appendix 1

Let's find conditions when such infinitive systems of coupled equations

$$\sum_{s=-N}^{N} \gamma_{n,n+s}^{(1,0)} A_{n+s} = 0 \,, \tag{1.15}$$

$$\sum_{s=-N}^{N} \gamma_{n,n+s}^{(2,0)} A_{n+s} = 0 \,, \tag{1.16}$$

have nonzero solutions. Using results of Appendix 2, we can rewrite (1.15) and (1.16) in the form

$$\begin{aligned} \sum_{s=N-1}^{N} \gamma_{n,n+s}^{(1,2N-1)} A_{n+s} = 0 \\ \sum_{s=N-1}^{N} \gamma_{n,n+s}^{(2,2N-1)} A_{n+s} = 0 \end{aligned} \,, \tag{1.17}$$

where $\gamma_{n,n+s}^{(1,2N-1)}$, $\gamma_{n,n+s}^{(2,2N-1)}$ are found from such recurrent relations :

$$p = 1 \qquad \gamma_{n,n+s}^{(1,1)} = \left( \gamma_{n,n-N}^{(1,0)} \gamma_{n,n-N}^{(2,0)} - \gamma_{n,n-N}^{(1,0)} \gamma_{n,n+s}^{(2,0)} \right) =$$

$$= \overline{\alpha}_{010}^{(n,n+s)} \overline{\alpha}_{010}^{(n,n-N)} \sin(\varphi_{n-N} - \varphi_{n+s})$$

$$\gamma_{n,n+s}^{(2,1)} = \left( \gamma_{n+1,n+s}^{(1,0)} \gamma_{n+1,n+1+N}^{(2,0)} - \gamma_{n+1,n+1+N}^{(1,0)} \gamma_{n+1,n+s}^{(2,0)} \right) = \quad , \tag{1.18}$$

$$= \overline{\alpha}_{010}^{(n+1,n+s)} \overline{\alpha}_{010}^{(n+1,n+1+N)} \sin(\varphi_{n+1+N} - \varphi_{n+s})$$

$$p = 2 \qquad \gamma_{n,n+s}^{(1,2)} = \left( \gamma_{n,n+s}^{(1,1)} \gamma_{n,n-N-1}^{(2,1)} - \gamma_{n,n-N-1}^{(1,1)} \gamma_{n,n+s}^{(2,1)} \right)$$

$$\gamma_{n,n+s}^{(2,2)} = \left( \gamma_{n+1,n+s}^{(1,1)} \gamma_{n+1,n+1+N}^{(2,1)} - \gamma_{n+1,n+1+N}^{(1,1)} \gamma_{n+1,n+s}^{(2,1)} \right) \quad , \tag{1.19}$$

$$p = 3 \qquad \gamma_{n,n+s}^{(1,3)} = \left( \gamma_{n,n+s}^{(1,2)} \gamma_{n,n-N-2}^{(2,2)} - \gamma_{n,n-N-2}^{(1,2)} \gamma_{n,n+s}^{(2,2)} \right)$$

$$\gamma_{n,n+s}^{(2,3)} = \left( \gamma_{n+1,n+s}^{(1,2)} \gamma_{n+1,n+1+N}^{(2,2)} - \gamma_{n+1,n+1+N}^{(1,1)} \gamma_{n+1,n+s}^{(2,2)} \right) \quad , \tag{1.20}$$

$$\ldots\ldots\ldots\ldots\ldots\ldots\ldots\ldots\ldots\ldots\ldots\ldots\ldots\ldots$$

$$p = 2N-1 \quad \gamma_{n,n+s}^{(1,2N-1)} = \left( \gamma_{n,n+s}^{(1,2N-2)} \gamma_{n,n+N-2}^{(2,2N-2)} - \gamma_{n,n+N-2}^{(1,2N-2)} \gamma_{n,n+s}^{(2,2N-2)} \right)$$

$$\gamma_{n,n+s}^{(2,2N-1)} = \left( \gamma_{n+1,n+s}^{(1,2N-2)} \gamma_{n+1,n+1+N}^{(2,2N-2)} - \gamma_{n+1,n+1+N}^{(1,2N-2)} \gamma_{n+1,n+s}^{(2,2N-2)} \right) \quad , \tag{1.21}$$

The system (1.17) has nonzero solutions if its determinant equals to zero

$$\gamma_{n,n+N-1}^{(1,2N-1)} \gamma_{n,n+N}^{(2,2N-1)} - \gamma_{n,n+N}^{(1,2N-1)} \gamma_{n,n+N-1}^{(2,2N-1)} = 0 . \tag{1.22}$$

In this case the infinitive systems of coupled equations (1.15) and (1.16) transform into simple ones

$$A_{n+N} = -\frac{\gamma_{n,n+N-1}^{(2,2N-1)}}{\gamma_{n,n+N}^{(2,2N-1)}} A_{n+N-1} = -\frac{\gamma_{n,n+N-1}^{(1,2N-1)}}{\gamma_{n,n+N}^{(1,2N-1)}} A_{n+N-1} , \tag{1.23}$$

Introducing $n + N = n'$ (and then changing $n'$ by $n$), we can rewrite (1.23) and (1.22) in the such form

$$A_n = -\frac{\gamma_{n-N,n-1}^{(2,2N-1)}}{\gamma_{n-N,n}^{(2,2N-1)}} A_{n-1} = -\frac{\gamma_{n-N,n-1}^{(1,2N-1)}}{\gamma_{n-N,n}^{(1,2N-1)}} A_{n-1} , \tag{1.24}$$

$$\gamma_{n-N,n-1}^{(1,2N-1)} \gamma_{n-N,n}^{(2,2N-1)} - \gamma_{n-N,n}^{(1,2N-1)} \gamma_{n-N,n-1}^{(2,2N-1)} = 0 . \tag{1.25}$$

## Appendix 2

Let's consider such infinitive systems of coupled equations

$$\sum_{s=-K}^{N} \beta_{n,n+s}^{(1)} A_{n+s} = 0 , \tag{1.26}$$

$$\sum_{s=-K}^{N} \beta_{n,n+s}^{(2)} A_{n+s} = 0 . \tag{1.27}$$

First we multiply the equation (1.26) by $\beta_{n,n-K}^{(2)}$, the equation (1.27) - by $\beta_{n,n-K}^{(1)}$ and subtract the results. We obtain

$$\sum_{s=-K+1}^{N} \left( \beta_{n,n+s}^{(1)} \beta_{n,n-K}^{(2)} - \beta_{n,n-K}^{(1)} \beta_{n,n+s}^{(2)} \right) A_{n+s} = 0 . \tag{1.28}$$

Multiplying the equation (1.26) by $\beta_{n,n+N}^{(2)}$, the equation (1.27) - by $\beta_{n,n+N}^{(1)}$ and subtracting the results, we obtain

$$\sum_{s=-K}^{N-1} \left( \beta_{n,n+s}^{(1)} \beta_{n,n+N}^{(2)} - \beta_{n,n+N}^{(1)} \beta_{n,n+s}^{(2)} \right) A_{n+s} = 0 \,. \tag{1.29}$$

If we increase the index $n$ in (1.29) by 1 ($n \Rightarrow n+1$), we can write

$$\sum_{s=-K}^{N-1} \left( \beta_{n+1,n+1+s}^{(1)} \beta_{n+1,n+1+N}^{(2)} - \beta_{n+1,n+1+N}^{(1)} \beta_{n+1,n+1+s}^{(2)} \right) A_{n+1+s} = 0 \,. \tag{1.30}$$

Introducing the new index $s' = s+1$ gives

$$\sum_{s'=-K+1}^{N} \left( \beta_{n+1,n+s'}^{(1)} \beta_{n+1,n+1+N}^{(2)} - \beta_{n+1,n+1+N}^{(1)} \beta_{n+1,n+s'}^{(2)} \right) A_{n+s'} = 0 \,. \tag{1.31}$$

Finally, we can write

$$\begin{aligned} \sum_{s=-K+1}^{N} \left( \beta_{n,n+s}^{(1)} \beta_{n,n-K}^{(2)} - \beta_{n,n-K}^{(1)} \beta_{n,n+s}^{(2)} \right) A_{n+s} = 0 \\ \sum_{s=-K+1}^{N} \left( \beta_{n+1,n+s}^{(1)} \beta_{n+1,n+1+N}^{(2)} - \beta_{n+1,n+1+N}^{(1)} \beta_{n+1,n+s}^{(2)} \right) A_{n+s} = 0 \end{aligned} \,. \tag{1.32}$$

In summery, we decreased the number of coupling elements by 1 in systems (1.26) and (1.27)

$$\begin{aligned} \sum_{s=-K}^{N} \beta_{n,n+s}^{(1)} A_{n+s} = 0 \Rightarrow \sum_{s=-K+1}^{N} \tilde{\beta}_{n,n+s}^{(1)} A_{n+s} = 0 \\ \sum_{s=-K}^{N} \beta_{n,n+s}^{(2)} A_{n+s} = 0 \Rightarrow \sum_{s=-K+1}^{N} \tilde{\beta}_{n,n+s}^{(2)} A_{n+s} = 0 \end{aligned} \,, \tag{1.33}$$

where

$$\begin{aligned} \tilde{\beta}_{n,n+s}^{(1)} = \left( \beta_{n,n+s}^{(1)} \beta_{n,n-K}^{(2)} - \beta_{n,n-K}^{(1)} \beta_{n,n+s}^{(2)} \right) \\ \tilde{\beta}_{n,n+s}^{(2)} = \left( \beta_{n+1,n+s}^{(1)} \beta_{n+1,n+1+N}^{(2)} - \beta_{n+1,n+1+N}^{(1)} \beta_{n+1,n+s}^{(2)} \right) \end{aligned} \,. \tag{1.34}$$